\documentclass[12pt]{article}

\usepackage{graphicx}
\begin{document}

\newcommand{\siml}{\stackrel{<}{\sim}}
\newcommand{\simg}{\stackrel{>}{\sim}}
\newcommand{\lleq}{\stackrel{<}{=}}


\begin{center}
{\large\bf
A moment approach to analytic time-dependent solutions \\
of the Fokker-Planck equation with \\
additive and multiplicative noise
} 
\end{center}

\baselineskip=1.333\baselineskip

\begin{center}
Hideo Hasegawa
\footnote{E-mail address: hideohasegawa@goo.jp}
\end{center}

\begin{center}
{\it Department of Physics, Tokyo Gakugei University,  \\
Koganei, Tokyo 184-8501, Japan}
\end{center}
\begin{center}
({\today})
\end{center}
\thispagestyle{myheadings}

\begin{abstract}
An efficient method is presented as a means 
of an approximate, analytic time-dependent solution
of the Fokker-Planck equation (FPE) for the Langevin model
subjected to additive and multiplicative noise.
We have assumed that the dynamical probability distribution 
function has the same
structure as the exact stationary one and that its parameters are expressed
in terms of first and second moments,
whose equations of motion are determined by the FPE.  
Model calculations have shown that
dynamical distributions in response to applied
signal and force calculated by
our moment method are in good agreement with
those obtained by the partial difference equation method.
As an application of our method, we present
the time-dependent Fisher information for the  
inverse-gamma distribution which is realized 
in the FPE including multiplicative noise only.

\end{abstract}

\vspace{0.5cm}


\newpage
\section{INTRODUCTION}

The Langvin model is a very important model to describe the 
diffusion behavior in non-equilibrium systems, and
it has been widely applied to various phenomena in
physics, chemistry and biology.
The Langevin model is usually transformed to the Fokker-Planck 
equation (FPE) which deals with the probability distribution
function (PDF) of a state variable \cite{Risken96}.
It is generally not possible to obtain analytic solutions
of the second-order partial equations.
Indeed, exact analytical solutions of the FPE are 
known for only a few cases.
In most cases, approximate solutions are obtained by 
using analytic or numerical methods.
Typical analytic methods are an appropriate change of variables,
eigenfunction expansion, perturbation expansion,
path integral, Green's function, moment method, and the 
continued-fraction method \cite{Risken96}. 
When no analytic solutions are available, numerical methods
such as finite-difference and finite-element methods 
have been employed.

For some Langevin models subjected to additive noise only, 
exact solutions have been obtained. For the linear Langevin model,
the exact dynamical solution is expressed by the Gaussian distribution  
with time-dependent mean and variance of a state variable.
For the FPE including a nonlinear diffusion term,
some authors have obtained exact dynamical solutions 
\cite{Borland99,Plastino00,Malacarne02}. 
The generalized FPEs in which time dependences
are introduced in drift and diffusion terms have been 
investigated \cite{Malacarne02}-\cite{Heinsale07}.

When multiplicative noise is incorporated to the Langevin model,
the problem becomes much difficult \cite{Munoz04}.
For the linear Langevin model subjected to
additive and multiplicative noise, 
the exact stationary solution is available, and it has been
considerably discussed in connection with the non-Gaussian
PDF in the nonextensive statistics 
\cite{Tsallis88}-\cite{Hasegawa07b}.
An exact dynamical solution for the linear Langevin model 
subjected to multiplicative noise only 
is obtained in Ref. \cite{Fa03} 
although it does not represent the stationary solution.
Approximate dynamical solutions of
the linear and nonlinear FPEs subjected to multiplicative noise
have been discussed with some sophisticated methods 
such as the polynomial expansion of the logarithmic PDF
\cite{Paola02}, the linearizing transformation \cite{Unal08} and 
the direct quadrature method for moment solution \cite{Attar08}.

Numerical methods are powerful approaches when exact dynamical 
solutions are not available. Analytical solutions are, however, 
indispensable in some subjects. 
A typical example is a calculation of the time-dependent
Fisher information which is expressed by the derivatives of the
dynamical PDF with respect to its parameters.
In a recent paper \cite{Hasegawa08a}, 
we calculated the Fisher information
in a typical nonextensive system described by the linear
Langevin model subjected to additive and multiplicative noise.
We developed an analytic dynamical approach to the FPE combined with
the $q$-moment method in which moments are
evaluated over the escort probability distribution \cite{Tsallis98}.
The dynamical PDFs calculated by our moment method
are shown to be in good agreement with those obtained by the
partial difference equation method (PDEM) \cite{Hasegawa08a}.
By using the calculated time-dependent PDF, we discussed
the dynamical properties of the Fisher information  
\cite{Hasegawa08a}. 
It is the purpose of the present study to extend such an analytical 
approach so as to be applied to a wide class of Langevin model
with the use of the conventional (normal) moment method instead of the
$q$-moment method.

The paper is organized as follows. In Sec. 2, we discuss the adopted 
Langevin model and moment method to obtain the dynamical PDF.
The developed method has been applied to the three Langevin models.
We present some numerical calculations of the time-dependent
PDF in response to an applied signal and force. 
Section 3 is devoted to conclusion and discussion on the dynamics of
Fisher information of the inverse-gamma distribution.

\section{METHOD AND RESULT}
\subsection{Fokker-Planck equation}

We have adopted the Langevin model subjected to 
cross-correlated additive ($\xi$)
and multiplicative noise ($\eta$) given by
\begin{eqnarray}
\frac{dx}{dt}\!\!&=&\!\! F(x) + G(x) \eta(t)
+ \xi(t)+I(t).
\label{eq:A1} 
\end{eqnarray}
Here 
$F(x)$ and $G(x)$ are arbitrary functions of $x$,
$I(t)$ stands for an external input, and $\eta(t)$ and $\xi(t)$ 
express zero-mean Gaussian white noises with correlations given by
\begin{eqnarray}
\langle \eta(t)\:\eta(t') \rangle
&=& \alpha^2 \:\delta(t-t'),\\
\langle \xi(t)\:\xi(t') \rangle 
&=& \beta^2 \: \delta(t-t'),\\
\langle \eta(t)\:\xi(t') \rangle &=& 
\epsilon \alpha \beta \: \delta(t-t'),
\label{eq:A2}
\end{eqnarray}
where $\alpha$ and $\beta$ denote the strengths of multiplicative
and additive noise, respectively, and $\epsilon$
the degree of the cross-correlation between the two noise.

The FPE is expressed by \cite{Tessone98,Liang04,Jin05}
\begin{eqnarray}
\frac{\partial}{\partial t}\: p( x,t) 
&=&- \frac{\partial}{\partial x}\left( \left[F(x) +I
+\left( \frac{\phi}{2} \right)
[\alpha^2 G(x)G'(x)+ \epsilon \alpha \beta\: G'(x)]
\right] \:p( x,t)\right)  
\nonumber \\
&+& \left(\frac{1}{2} \right) \frac{\partial^2}{\partial x^2} 
\{[\alpha^2 G(x)^2+ 2 \epsilon \alpha\beta G(x)+\beta^2]\:p(x,t) \},
\label{eq:A3}
\end{eqnarray}
where $G'(x)=dG(x)/dx$, 
and $\phi=0$ and 1 in the Ito and Stratonovich
representations, respectively. 
Although we have adopted the Langevin model for a single
variable in this study, it is straightforward to extend it to the
coupled Langevin model with the use of the mean-field
approximation \cite{Hasegawa08a}.

For $I(t)=I$, the stationary PDF of $p(x)$
is expressed by \cite{Hasegawa07c}
\begin{eqnarray}
\ln p(x) &=& X(x)+Y(x)
-\left(1- \frac{\phi}{2} \right)
\ln \left(\frac{1}{2}
[\alpha^2 G(x)^2 +2\epsilon \alpha \beta G(x)+ \beta^2] \right),
\label{eq:A4}
\end{eqnarray}
with
\begin{eqnarray}
X(x) &=& 2 \int \:dx \:
\left[ \frac{F(x)}{\alpha^2 G(x)^2+2\epsilon \alpha \beta G(x)
+\beta^2} \right], \\
Y(x) &=& 2 \int \:dx \:
\left[ \frac{I}{\alpha^2 G(x)^2+2\epsilon \alpha \beta G(x)
+\beta^2} \right].
\label{eq:A5}
\end{eqnarray}

\subsection{Equations of motion for the moments}

An equation of motion for the $n$th moment is given by
\begin{eqnarray}
\frac{\partial \langle x^n \rangle}{\partial t} 
&=& \int \frac{\partial p(x,t)}{\partial t}\:x^n \:dx, \\
&=& n \left( \left< x^{n-1} F(x) \right>+ \left< x^{n-1} I(t) \right>
+\frac{\phi}{2}\left[\alpha^2\left< x^{n-1} G(x)G'(x) \right> 
+ \epsilon \alpha \beta \left< x^{n-1} G'(x) \right>\right] \right)\nonumber \\
&+& \frac{n(n-1)}{2}\left [\alpha^2 \left< x^{n-2}G(x)^2 \right>
+ 2 \epsilon \alpha \beta \left< x^{n-2}G(x) \right>
+ \beta^2 \left< x^{n-2} \right> \right],
\label{eq:A10}
\end{eqnarray}
where suitable boundary conditions are adopted.
For $n=1,\:2$, we obtain 
\begin{eqnarray}
\frac{\partial \langle x \rangle}{\partial t}
&=& \langle F(x) \rangle + \langle I(t) \rangle
+ \frac{\phi}{2} [\alpha^2 \langle G(x) G'(x) \rangle 
+ \epsilon \alpha \beta \langle G'(x) \rangle],
\label{eq:A6} \\
\frac{\partial \langle x^2 \rangle}{\partial t}
&=& 2 \langle x F(x) \rangle+  2 \langle x I(t) \rangle
+ \phi [\alpha^2 \langle x G(x) G'(x) \rangle 
+ \epsilon \alpha \beta \langle x G'(x) \rangle] \nonumber \\
&+& \alpha^2 \langle G(x)^2 \rangle 
+ 2 \epsilon \alpha \beta \langle G(x) \rangle +\beta^2.
\label{eq:A7}
\end{eqnarray}
Expanding $x$ as $x=\mu+\delta x$ and retaining up to
$O(\langle (\delta x)^2 \rangle)$, we obtain
equations of motion for the
average $\mu$ [$=\langle x \rangle $] and
variance $\sigma^2$ [$=\langle x^2 \rangle -\langle x \rangle^2$] 
given by \cite{Hasegawa07b}
\begin{eqnarray}
\frac{d \mu}{dt}&=& f_0+f_2\sigma^2 
+\frac{\phi }{2}
\left( \alpha^2[g_0g_1+3(g_1g_2+g_0g_3)\sigma^2]
+ \epsilon \alpha \beta(g_1+3 g_3 \sigma^2) \right)+I(t), 
\label{eq:A8}\\
\frac{d \sigma^2}{dt} &=& 2f_1 \sigma^2 
+ (\phi+1) (g_1^2+2 g_0g_2)\alpha^2 \sigma^2
+2 \epsilon \alpha \beta (\phi+1) g_2 \sigma^2 \nonumber \\
&+& \alpha^2 g_0^2+ 2 \epsilon \alpha \beta g_0 +\beta^2,
\label{eq:A9}
\end{eqnarray}
where $f_{\ell}=(1/\ell !)
\partial^{\ell} F(\mu)/\partial x^{\ell}$ and
$g_{\ell}=(1/\ell !) 
\partial^{\ell} G(\mu)/\partial x^{\ell}$.

\subsection{Model A}

\subsubsection{Stationary distribution}

Our dynamical moment approach 
will be applied to the three Langevin models
A, B and C, which will be separately discussed 
in Secs. 2,3, 2.4 and 2.5, respectively. 

First we consider the model A in which $F(x)$ and $G(x)$ are given by
\begin{eqnarray}
F(x) &=& -\lambda x, 
\label{eq:B0} \\
G(x) &=& x,
\label{eq:B1}
\end{eqnarray}
with $\epsilon=0.0$ ({\it i.e.,} without the cross-correlation),
where $\lambda $ expresses the relaxation rate.
The model A has been adopted as a microscopic model
for nonextensive systems \cite{Sakaguchi01}-\cite{Hasegawa07b}. 
From Eq. (\ref{eq:A3}), the FPE in the Stratonovich representation 
is given by
\begin{eqnarray}
\frac{\partial}{\partial t}\: p(x,t) 
&=& \frac{\partial}{\partial x} \left[\lambda x -I(t)\right] p(x,t) 
+ \left( \frac{\beta^2}{2} \right)
\frac{\partial^2}{\partial x^2}p(x,t) \nonumber
\\
&+& \left( \frac{\alpha^2}{2} \right)
\frac{\partial}{\partial x} 
\left[ x \frac{\partial}{\partial x} \{ x p(x,t) \} \right].
\label{eq:B2}
\end{eqnarray}
By using Eqs. (\ref{eq:A4})-(\ref{eq:A5}),
we obtain the stationary PDF given by \cite{Hasegawa08a}
\begin{eqnarray}
p(x) &=& \left( \frac{1}{Z} \right)
\frac{\exp [2c \tan^{-1}(ax)]}{(1+a^2x^2)^b},
\label{eq:B3}
\end{eqnarray}
with
\begin{eqnarray}
a &=& \frac{\alpha}{\beta}, 
\label{eq:B4} \\
b &=& \frac{(2\lambda+\alpha^2)}{2 \alpha^2}, 
\label{eq:B5} \\
c &=& \frac{I}{\alpha \beta}, 
\label{eq:B6} \\
Z &=& \frac{\sqrt{\pi}\:\Gamma(b)\Gamma(b-\frac{1}{2})}
{a\:\mid \Gamma(b+ i c) \mid^2}.
\label{eq:B7}
\end{eqnarray}
By using Eq. (\ref{eq:B3}), we obtain the mean and variance in
the stationary state given by 
\begin{eqnarray}
\mu &=& \frac{c}{a (b-1)}=\frac{2I}{(2 \lambda-\alpha^2)}, 
\label{eq:B8} \\
\sigma^2 &=& \frac{[(b-1)^2+c^2]}{a^2(b-1)^2(2b-3)}
=\frac{(\alpha^2 \mu^2+\beta^2)}{2(\lambda-\alpha^2)}.
\label{eq:B9}
\end{eqnarray}
Depending on the model parameters, the stationary PDF 
given by Eq. (\ref{eq:B3}) may reproduce various 
PDFs such as the Gaussian, $q$-Gaussian, 
Cauchy and inverse-gamma PDFs \cite{Hasegawa08a}.

\subsubsection{Dynamical distribution}

It is worthwhile to remind the dynamical solution of the FPE 
given by Eq. (\ref{eq:B2}) in the limit of $\alpha = 0.0$
({\it i.e.,} additive noise only), for which
the time-dependent solution is given by
\begin{equation}
p(x,t)= \frac{1}{\sqrt{2 \pi \:\sigma(t)^2}}
\;e^{-[x-\mu(t)]^2/2 \sigma(t)^2},
\label{eq:C1}
\end{equation}
with $\mu(t)$ and $\sigma(t)^2$ satisfying equations of motion given by
\begin{eqnarray}
\frac{d \mu(t)}{dt} &=& -\lambda \mu(t)+ I(t), 
\label{eq:C2} 
\\
\frac{d \sigma(t)^2}{dt} &=& -2 \lambda \sigma(t)^2 + \beta^2.
\label{eq:C3}
\end{eqnarray}

In order to derive the dynamical solution of the FPE
for $\alpha \neq 0.0$ given by Eq. (\ref{eq:B2}), 
we adopt the moment approach with the following steps:

\noindent
(1) We assume that dynamical PDF has the
same structure as the stationary one, as given by
\begin{eqnarray}
p(x,t) &=& \left( \frac{1}{Z(t)} \right)
\frac{ \exp [2 c(t) \:\tan^{-1} \{a(t) x\} ]  }
{ [1+ a(t)^2 x^2 ]^{b(t)}},
\label{eq:C4}
\end{eqnarray}
with
\begin{eqnarray}
Z(t) &=& \frac{\sqrt{\pi}\:\Gamma[b(t)]\:\Gamma[b(t)-\frac{1}{2}]}
{a(t)\:\mid \Gamma[b(t)+ i c(t)] \mid^2}.
\label{eq:C5}
\end{eqnarray}

\noindent
(2) With the assumption (1), we first tried to derive equations of motion
for the parameters of $a(t)$, $b(t)$ and $c(t)$, by using the FPE after 
Refs. \cite{Borland99,Plastino00,Malacarne02}. Unfortunately, it did
not work because functional forms 
in the left and right sides of the FPE become different.

Then we tried to express the parameters 
in terms of importance quantities of $\mu(t)$ and $\sigma(t)^2$
such as to be consistent with the relations for
the stationary state given by 
Eqs. (\ref{eq:B8}) and (\ref{eq:B9}).
Because the number of parameters (three)
is larger than two for $\mu(t)$ and $\sigma(t)^2$,
the parameters of $a(t)$, $b(t)$ and $c(t)$ cannot be uniquely
expressed in terms of $\mu(t)$
and $\sigma(t)^2$ from Eqs. (\ref{eq:B8}) and (\ref{eq:B9}).
If the first three moments in the stationary state are available, 
it is possible to uniquely express $a(t)$, $b(t)$ and $c(t)$
in terms of them, though such a calculation is laborious.

In order to overcome the above problem,
we have imposed an additional condition that
expressions for the parameters 
yield the consistent result in the two limiting cases of
$\alpha \rightarrow 0$ and $\beta \rightarrow 0$.
After several tries, we have decided that $b(t)$ and $c(t)$ 
in Eqs. (\ref{eq:C4}) and (\ref{eq:C5}) are expressed as  
\begin{eqnarray}
b(t) &=& \frac{[1+a^2\{\mu(t)^2+3 \sigma(t)^2\}]}
{2 a^2 \sigma(t)^2}, 
\label{eq:C6}\\
c(t) &=& \frac{[1+a^2\{\mu(t)^2+\sigma(t)^2\}] \mu(t)}
{2 a \:\sigma(t)^2},
\label{eq:C7}
\end{eqnarray}
with the time-independent $a$ 
($=\alpha/\beta$) given by Eq. (\ref{eq:B4}).
The relations given by Eqs. (\ref{eq:C6}) and (\ref{eq:C7}) 
are consistent with Eqs. (\ref{eq:B8}) and (\ref{eq:B9})
for the stationary state and they satisfy the above-mentioned 
limiting conditions, as will be shown shortly.

\noindent
(3) Equations of motion for $\mu(t)$ and $\sigma(t)^2$ 
in Eqs. (\ref{eq:C6}) and (\ref{eq:C7}) are obtained from 
Eqs. (\ref{eq:A8})-(\ref{eq:B1}), 
as given by
\begin{eqnarray}
\frac{d \mu(t)}{dt}&=&-\lambda \mu(t) + I(t)
+ \frac{\alpha^2 \mu(t)}{2}, 
\label{eq:C8}\\
\frac{d \sigma(t)^2}{dt} &=& -2 \lambda \sigma(t)^2 
+ 2 \alpha^2 \sigma(t)^2 + \alpha^2 \mu(t)^2 + \beta^2.
\label{eq:C9}
\end{eqnarray}
Thus the dynamical solution of the FPE given by Eq. (\ref{eq:B2})
is expressed by Eqs. (\ref{eq:C4})-(\ref{eq:C9}).

In the following, we will show that
the relations given by Eqs. (\ref{eq:C6}) and (\ref{eq:C7})
lead to results consistent in the two limiting cases of 
$\alpha \rightarrow 0.0$ and $\beta \rightarrow 0.0$.

\vspace{0.5cm}
\noindent
{\bf (a) $\alpha \rightarrow 0$ case}

In the limit of $\alpha \rightarrow 0.0$ 
({\it i.e.,} additive noise only),
$p(x,t)$ given by Eq. (\ref{eq:C4}) reduces to
\begin{eqnarray}
p(x,t) &\propto& e^{-a(t)^2b(t) x^2+2a(t)c(t) x} \\
&\rightarrow & e^{-[x-\mu(t)]^2/2\sigma(t)^2},
\label{eq:D3}
\end{eqnarray}
because Eqs. (\ref{eq:C6}) and (\ref{eq:C7}) 
with $a \rightarrow 0.0$ yield
\begin{eqnarray}
a(t)^2 b(t) &=& \frac{[1+a^2\{\mu(t)^2+3 \sigma(t)^2\}]}{2 \sigma(t)^2}
\rightarrow \frac{1}{2 \sigma(t)^2}, \\
2a(t)c(t) & = & 
\frac{[1+a^2\{\mu(t)^2+\sigma(t)^2\}]\:\mu(t)}{\sigma(t)^2}
\rightarrow \frac{\mu(t)}{\sigma(t)^2}. 
\end{eqnarray}
Equation (\ref{eq:D3}) agrees with the Gaussian distribution
given by Eq. (\ref{eq:C1})

\vspace{0.5cm}
\noindent
{\bf (b) $\beta \rightarrow 0$ case}

In the opposite limit of $\beta=0.0$
({\it i.e.,} multiplicative noise only), the stationary PDF
given by Eqs. (\ref{eq:B3}) and (\ref{eq:B7}) with $I > 0$ 
leads to the inverse-gamma distribution expressed by
\begin{eqnarray}
p(x) &=& \frac{\kappa^{\delta-1}}{\Gamma[\delta-1]}
\:x^{-\delta} e^{-\kappa/x}\: \Theta(x),
\label{eq:E1} 
\end{eqnarray}
where
\begin{eqnarray}
\delta &=& 2b, 
\label{eq:E2} \\
\kappa &=& \frac{2c}{a}.
\label{eq:E3}
\end{eqnarray}
Here $\Gamma(x)$ denotes the gamma function and
$\Theta(t)$ the Heaviside 
function: $\Theta(t)=1$ for $t > 0$ and zero otherwise.
From Eq. (\ref{eq:E1}), we obtain
the average and variance in the stationary state given by
\begin{eqnarray}
\mu &=&  \frac{\kappa}{(\delta-2)}, 
\label{eq:E4}\\
\sigma^2 &=& \frac{\kappa^2}{(\delta-2)^2 (\delta-3)},
\label{eq:E5}
\end{eqnarray}
from which $\delta$ and $\kappa$ are expressed
in terms of $\mu$ and $\sigma^2$ as
\begin{eqnarray}
\delta &=& \frac{\mu^2+3 \sigma^2}{\sigma^2}, 
\label{eq:E6}\\
\kappa &=& \frac{\mu^2+\sigma^2}{\sigma^2}\:\mu.
\label{eq:E7}
\end{eqnarray}

On the contrary, the dynamical PDF given by 
Eq. (\ref{eq:C4}) in the limit of $\beta \rightarrow 0.0$ 
(and $I > 0 $) reduces to
\begin{eqnarray}
p(x, t) \propto \frac{e^{-2c(t)/a(t)x}}{x^{2b(t)}}\:\Theta(x)
\rightarrow \frac{e^{-\kappa(t)/x}}{x^{\delta(t)}}\:\Theta(x),
\label{eq:E8}
\end{eqnarray}
because Eqs. (\ref{eq:C6}) and (\ref{eq:C7}) with
$\beta \rightarrow 0.0$ ($a \rightarrow \infty$) lead to
\begin{eqnarray}
\delta(t) &=& 2b(t)
= \frac{[1+a^2\{\mu(t)^2+3 \sigma(t)^2\}]}{a^2\sigma(t)^2}
\rightarrow \frac{\mu(t)^2+3\sigma(t)^2}{\sigma(t)^2}, 
\label{eq:E9} \\
\kappa(t) &=& \frac{2c(t)}{a(t)}
=\frac{[1+a^2\{\mu(t)^2+\sigma(t)^2\}]\mu(t)}{a^2\sigma(t)^2}
\rightarrow \frac{\mu(t)^2+\sigma(t)^2}{\sigma(t)^2}\:\mu(t).
\label{eq:E10} 
\end{eqnarray}
Equations (\ref{eq:E8}), (\ref{eq:E9}) and (\ref{eq:E10}) agree 
with Eqs. (\ref{eq:E1}), (\ref{eq:E6}) and (\ref{eq:E7}), respectively.

Thus the expressions given by Eqs. (\ref{eq:C6}) and (\ref{eq:C7}) 
yield the consistent result
covering the two limits of $\alpha \rightarrow 0.0$ and 
$\beta \rightarrow 0.0$.

\subsection{Model B}

\subsubsection{Stationary distribution}

Next we consider the model B in which $F(x)$ and $G(x)$ are given 
by Eqs. (\ref{eq:B0}) and (\ref{eq:B1}) with $\epsilon \neq 0$. 
From Eqs. (\ref{eq:A4})-(\ref{eq:A5}), the stationary PDF 
is given by 
\begin{eqnarray}
p(x)&=& \left( \frac{1}{Z} \right)
\frac{ \exp[2c \tan^{-1} \{a(x+f)\} ] } 
{[1+a^2(x+f)^2]^b}, 
\label{eq:F1}
\end{eqnarray}
with
\begin{eqnarray}
a &=& \frac{\alpha}{\beta \sqrt{1-\epsilon^2}}, 
\label{eq:F2} \\
b &=& \frac{2\lambda+\alpha^2}{2 \alpha^2}, 
\label{eq:F3}\\
c &=& \frac{(I+\lambda f)}
{\alpha  \beta \sqrt{1-\epsilon^2}}, 
\label{eq:F4}\\
f &=& \frac{\epsilon \beta}{\alpha}, 
\label{eq:F5}\\
Z &=& \frac{\sqrt{\pi}\:\Gamma(b)\Gamma(b-\frac{1}{2})}
{a\:\mid \Gamma(b+ i c) \mid^2}.
\label{eq:F6}
\end{eqnarray}
Equations (\ref{eq:F1}) and (\ref{eq:F6}) yield the average 
and variance in the stationary state given by
\begin{eqnarray}
\mu &=& \frac{c}{a(b-1)}-f
= \frac{(2 I +\epsilon \alpha \beta)}{(2 \lambda-\alpha^2)},
\label{eq:F7} \\
\sigma^2 &=& \frac{[(b-1)^2+c^2]}{a^2(b-1)^2(2b-3)} 
= \frac{(\alpha^2 \mu^2+2 \epsilon \alpha\beta \mu+\beta^2)}
{2(\lambda-\alpha^2)}.
\label{eq:F8}
\end{eqnarray}

\subsubsection{Dynamical distribution}

With the use of the procedure mentioned for the model A in Sec. 2.3,
the dynamical PDF $p(x,t)$
of the model B is assumed to be
given by Eq. (\ref{eq:F1}) but with $b$ and $c$ replaced by
\begin{eqnarray}
b(t) &=& \frac{[1+a^2\{[\mu(t)+f]^2+3 \sigma(t)^2\}]}
{2 a^2 \sigma(t)^2}, 
\label{eq:G1} \\
c(t) &=& \frac{[1+a^2\{[\mu(t)+f]^2+\sigma(t)^2\}] [\mu(t)+f]}
{2 a \sigma(t)^2},
\label{eq:G2}
\end{eqnarray}
which agree with Eqs. (\ref{eq:F7}) and (\ref{eq:F8})
in the stationary state.
Equations of motion for $\mu(t)$ and $\sigma(t)^2$
in Eqs. (\ref{eq:G1}) and (\ref{eq:G2}) are given by
\begin{eqnarray}
\frac{d \mu(t)}{dt}&=&-\lambda \mu(t) + I(t)
+ \frac{\alpha^2 \mu(t)}{2}+ \frac{\epsilon \alpha \beta}{2}, 
\label{eq:G3}\\
\frac{d \sigma(t)^2}{dt} &=& -2 \lambda \sigma(t)^2 
+ 2 \alpha^2 \sigma(t)^2 
+ \alpha^2 \mu(t)^2 + 2 \epsilon \alpha \beta \mu(t) + \beta^2,
\label{eq:G4}
\end{eqnarray}
which are derived from Eqs. (\ref{eq:A8})-(\ref{eq:B1}).

It is noted that in the limits of $\alpha \rightarrow 0.0$
and $\beta \rightarrow 0.0$, the cross-correlation
between additive and multiplicative noise
does not work, and the result for the model B reduces to that
for the model A.
Thus the moment method
with the use of Eqs. (\ref{eq:G1}) and (\ref{eq:G2}) 
leads to the results consistent in the limits of $\alpha \rightarrow 0.0$
and $\beta \rightarrow 0.0$, where $p(x,t)$ becomes
the Gaussian and inverse-gamma distributions, respectively.

\subsection{Model C}

\subsubsection{Stationary distribution}

Now we consider the model C in which $F(x)$ and $G(x)$ are given by
\begin{eqnarray}
F(x) &=& -\lambda (x+s),
\label{eq:H1} \\
G(x) &=& \sqrt{x^2+ 2 sx+r^2 },
\label{eq:H2}
\end{eqnarray}
with $\epsilon=0.0$ where $\lambda $ expresses the relaxation rate,
and $r$ and $s$ are parameters.
We assume that $\epsilon=0.0$ because we cannot obtain
the analytic stationary PDF for $\epsilon \neq 0.0$.
The model C with $r=s=0.0$ is nothing but the model A.
From Eqs. (\ref{eq:A4})-(\ref{eq:A5}), 
the stationary PDF for the model C is given by
\begin{eqnarray}
p(x) & \propto & 
\left[1-\frac{\alpha^2}{D}(x+s)^2 \right]^{-b}\:e^{Y(x)},
\label{eq:H3}
\end{eqnarray}
where
\begin{eqnarray}
Y(x) &=& \left( \frac{I}{\alpha \sqrt{D}} \right)
\ln \left| \frac{x+s-\sqrt{D}}{x+s+\sqrt{D}} \right|,
\hspace{1cm}\mbox{for $D > 0$} 
\label{eq:H4} \\
&=& \left( \frac{2I}{\alpha \sqrt{-D}} \right)
\tan^{-1} \left( \frac{\alpha(x+s)}{\sqrt{-D}} \right),
\hspace{1cm}\mbox{for $D < 0$} 
\label{eq:H5}\\
&=& - \frac{2I}{\alpha^2 (x+s)},
\label{eq:H6}
\hspace{3cm}\mbox{for $D = 0$}
\end{eqnarray}
with 
\begin{eqnarray}
D &=& \alpha^2(s^2-r^2)-\beta^2,
\label{eq:H7} \\
b &=& \frac{(2\lambda+\alpha^2)}{2 \alpha^2}.
\label{eq:H8}
\end{eqnarray}
When we consider the case of $D < 0$,
the stationary PDF is rewritten as
\begin{eqnarray}
p(x) &=& \left( \frac{1}{Z} \right)
\frac{ \exp[2c \tan^{-1} \{a(x+s)\} ] }
{[1+a^2(x+s)^2]^b},
\label{eq:H9}
\end{eqnarray}
with
\begin{eqnarray}
a &=& \frac{\alpha}{\sqrt{\beta^2+\alpha^2(r^2-s^2)}},
\label{eq:H10} \\
c &=& \frac{I}{\alpha \sqrt{\beta^2+\alpha^2(r^2-s^2)} }, \\
\label{eq:H11} 
Z &=& \frac{\sqrt{\pi}\:\Gamma(b)\Gamma(b-\frac{1}{2})}
{a\:\mid \Gamma(b+ i c) \mid^2}.
\label{eq:H12}
\end{eqnarray}
From Eq. (\ref{eq:H9}), we obtain $\mu$ and $\sigma^2$ in
the stationary state expressed by
\begin{eqnarray}
\mu &=& \frac{c}{a(b-1)}-s
=\frac{2I}{(2\lambda-\alpha^2)}-s,
\label{eq:H13} \\
\sigma^2 &=& \frac{[(b-1)^2+c^2]}{a^2 (b-1)^2(2b-3)}
=\frac{[\alpha^2 (\mu^2+2s\mu+r^2) + \beta^2]}
{2(\lambda-\alpha^2)}.
\label{eq:H14}
\end{eqnarray}

\subsubsection{Dynamical distribution}

In order to obtain the dynamical PDF for the model C,
we adopt the same procedure 
as those for the models A and B.
We assume that the dynamical solution $p(x,t)$ of the model C 
is given by Eq. (\ref{eq:H9}) but with $b$ and $c$ replaced by
\begin{eqnarray}
b(t) &=& \frac{[1+a^2\{[\mu(t)+s]^2+3 \sigma(t)^2\}]}
{2 a^2 \sigma(t)^2},
\label{eq:J1} \\
c(t) &=& \frac{[1+a^2\{[\mu(t)+s]^2+\sigma(t)^2\}] [\mu(t)+s]}
{2 a \sigma(t)^2},
\label{eq:J2}
\end{eqnarray}
which agree with Eqs. (\ref{eq:H13}) and (\ref{eq:H14})
in the stationary state.
Equations of motion for
$\mu(t)$ and $\sigma(t)^2$ in Eqs. (\ref{eq:J1}) and (\ref{eq:J2})
are given by
\begin{eqnarray}
\frac{d \mu(t)}{dt}&=& -\left(\lambda -\frac{\alpha^2}{2} \right)
[\mu(t)+s] + I(t),
\label{eq:J3} \\  
\frac{d \sigma(t)^2}{dt} &=& -2 (\lambda-\alpha^2) \sigma(t)^2 
+ \alpha^2 [\mu(t)^2+2s\mu(t)+r^2]+\beta^2, 
\label{eq:J4}
\end{eqnarray}
which are derived from Eqs. (\ref{eq:A8}), (\ref{eq:A9}),
(\ref{eq:H1}) and (\ref{eq:H2}).
It is easy to see that the moment method
with the use of Eqs. (\ref{eq:J1}) and (\ref{eq:J2}) 
lead to the results consistent in the limits of $\alpha \rightarrow 0.0$
and $\beta \rightarrow 0.0$, where $p(x,t)$ becomes
the Gaussian and inverse-gamma distributions, respectively.

\subsection{Model calculations}

We will present some numerical calculations in this subsection.
In order to examine the validity of the moment
approach, we have employed the partial difference
equation derived from Eq. (\ref{eq:A3})
with $\phi=1$, as given by 
\begin{eqnarray}
p(x,t+v) &=& p(x,t) 
+\left(-F'
+ \frac{\alpha^2}{2}[(G')^2+GG^{(2)}]
+ \frac{\epsilon \alpha \beta}{2} G^{(2)} \right) v \:p(x,t) \nonumber\\
&+&\left[ -F-I(t)
+ \frac{3 \alpha^2}{2}  G G' 
+ \frac{3 \epsilon \alpha \beta}{2}G' \right]
\left(\frac{v}{2 u}\right)[p(x+u)-p(x-u)] \nonumber \\
&+& \left(\frac{\alpha^2}{2} G^2 + \epsilon \alpha \beta G
+ \frac{\beta^2}{2} \right)
\left(\frac{v}{u^2}\right)[p(x+u,t)+p(x-u,t)-2p(x,t)],
\label{eq:L1}
\end{eqnarray}
where $u$ and $v$ denote incremental steps of $x$ and $t$, respectively.
We impose the boundary condition:
\begin{eqnarray}
p(x,t)=0, \hspace{1cm}\mbox{for $ \mid x \mid \ge x_m$}
\label{eq:L2}
\end{eqnarray}
with $x_m=5$, and the initial condition of $p(x,0)=p_0(x)$ where $p_0(x)$ 
is the stationary PDF.
We have chosen parameters of $u=0.05$ and $v=0.0001$ such as to satisfy 
the condition: $(\alpha^2 x_m^2 v/2 u^2) < 1/2$, which is required for
stable, convergent solutions of the PDEM.

First we apply a pulse input signal given by
\begin{equation}
I(t) = \Delta I \:\Theta(t-2)\Theta(6-t)+I_b,
\label{eq:M1}
\end{equation}
with $\Delta I=0.5$ and $I_b=0.0$ 
to the model B with $\lambda=1.0$, $\alpha=0.5$,
$\beta=0.5$ and $\epsilon=0.5$.
Figure \ref{figK} shows the time-dependence of the 
PDF at various $t$ in response to an applied input.
Solid curves express the results of the moment method 
calculated with the use of 
Eqs. (\ref{eq:F1}), (\ref{eq:F6}), (\ref{eq:G1})-(\ref{eq:G4}), 
and dashed curves denote those of the PDEM 
with Eq. (\ref{eq:L1}).
Figure \ref{figJ} shows the time-dependence
of $\mu(t)$ and $\sigma(t)^2$ calculated by the moment method 
with Eqs. (\ref{eq:G3}) and (\ref{eq:G4}) (solid curves)
and by the PDEM with Eq. (\ref{eq:L1}) (dashed curves).
At $0 \leq t < 2.0$ where no input signal is applied, 
we obtain $\mu(t)=0.071$ and $\sigma(t)^2=0.179$.
The PDF at $ t < 2.0$ is not symmetric with respect to its
center because of the introduced correlation of $\epsilon=0.5$.
By an applied pulse at $2.0 \leq t < 6.0$, 
$\mu(t)$ and $\sigma(t)^2$ are increased, and
the position of $p(x,t)$ moves rightward 
with slightly distorted shapes.
After an input pulse diminishes at $t \geq 6.0$, the PDF
gradually restores to its original stationary shape.

Next we apply the pulse input given by Eq. (\ref{eq:M1})
with $\Delta I=0.5$ and $I_b=0.0$  
to the model C with $\lambda=1.0$, $\alpha=0.5$,
$\beta=0.5$, $r=0.5$ and $s=0.2$.
Figure \ref{figB} shows the time-dependent 
PDF at various $t$ in response to an applied input.
Results of the moment method 
calculated with the use of 
Eqs. (\ref{eq:H9}), (\ref{eq:H12}), (\ref{eq:J1})-(\ref{eq:J4})
are shown by solid curves while 
those of the PDEM with Eq. (\ref{eq:L1})
are expressed by dashed curves. 
The time-dependent $\mu(t)$ and $\sigma(t)^2$ are shown in
Fig. \ref{figA}: solid curves denote the results calculated 
by the moment method with Eqs. (\ref{eq:J3}) and (\ref{eq:J4}): 
dashed curves express those by the PDEM with Eq. (\ref{eq:L1}).
In the stationary state at $t < 2.0$, 
we obtain $\mu(t)=-0.2$ and $\sigma(t)^2=0.2$.
By an applied pulse at $2.0 \leq t < 6.0$, 
$\mu(t)$ and $\sigma(t)^2$ are increased and
the position of $p(x,t)$ moves rightward 
with slightly changed shapes.

It is possible to calculate the response to temporal changes
in the model parameters such as $\lambda$, $\alpha$ and $\beta$. 
As an example, we introduce the time-dependent
relaxation rate $\lambda(t)$ given by
\begin{equation}
\lambda(t) = \Delta\lambda \:\Theta(t-2)\Theta(6-t)+\lambda_b,
\label{eq:M2}
\end{equation}
with $\Delta \lambda=1.0$ and $\lambda_b=1.0$
to the model C with $\alpha=0.5$, $\beta=0.0$,
$r=0.5$, $s=0.2$ and $I=0.0$.
Equation (\ref{eq:M2}) stands for an application of
an external force of $-\Delta \lambda \:x$
at $2.0 \leq t < 6.0$.
The time dependent $p(x,t)$ is plotted in Fig. \ref{figF},
where solid and dashed curves
denote the results of the moment method and PDEM, respectively.
Figure \ref{figG} shows the time dependences of
$\mu(t)$ and $\sigma(t)^2$.
When an external force is applied at $2.0 \leq t < 6.0$, 
the width of the PDF is reduced and
$\sigma(t)^2$ is decreased while $\mu(t)$ has no changes.

It is noted in Figs. \ref{figK}-\ref{figG} that
results of $p(x, t)$, $\mu(t)$ and $\sigma(t)^2$
calculated by the moment method are in good 
agreement with those obtained by the PDEM.

\section{CONCLUSION AND DISCUSSION}

The moment approach to the FPE 
discussed in preceding Sec. 2
may be applied to various Langevin models provided
analytic expressions for the PDF and for  
the first- and second-order moments in the
stationary state are available.
For example, when $F(x)$ and $G(x)$ are given by
\begin{eqnarray}
F(x) &=& -\lambda x \vert x \vert^{r-1}, 
\label{eq:P1} \\
G(x) &=& x \vert x \vert^{s-1}, 
\label{eq:P2}
\end{eqnarray}
for $r \geq 0$, $s \geq 0$,
the stationary PDF with $\epsilon=0.0$
is given by \cite{Anten02,Hasegawa07c}
\begin{eqnarray}
p(x) \propto (\alpha^2 \vert x \vert^{2s}+\beta^2)^{-1/2}
\exp[X(x)+Y(x)],
\label{eq:P3}
\end{eqnarray}
with
\begin{eqnarray}
X(x) &=& -\left( \frac{2\lambda \vert x \vert^{r+1}}{\beta^2 (r+1)}\right)
F\left(1, \frac{r+1}{2s}, \frac{r+1}{2s}+1;
-\frac{\alpha^2 \vert x \vert^{2s}}{\beta^2} \right), \\
Y(x) &=& \left( \frac{2 I \vert x \vert }{\beta^2}\right)
F\left(1, \frac{1}{2s}, \frac{1}{2s}+1;
-\frac{\alpha^2 \vert x \vert^{2s}}{\beta^2} \right),
\label{eq:P4} 
\end{eqnarray}
where $F(a,b,c;z)$ denotes the hypergeometric function.
Equations of motion for $\mu$ and $\sigma^2$ are given by
\cite{Hasegawa07b}
\begin{eqnarray}
\frac{d\mu}{dt} &=& -\lambda \mu \vert \mu\vert^{r-1}+I 
-\left( \frac{\lambda}{2} \right) r(r-1)
\mu \vert \mu\vert^{r-3}\sigma^2 \nonumber \\
&+& \left( \frac{\alpha^2}{2}\right)
[s \mu \vert \mu \vert^{2s-2}
+s(s-1)(2s-1)\mu \vert \mu\vert ^{2s-4}\sigma^2], 
\label{eq:P5} \\
\frac{d \sigma^2}{dt} &=& -2 \lambda r \vert \mu \vert^{r-1} \sigma^2
+2s(2s-1)\alpha^2 \vert \mu \vert^{2s-2} \sigma^2
+\alpha^2 \vert \mu \vert^{2s}+\beta^2.
\label{eq:P6}
\end{eqnarray} 
If analytic expressions for stationary values 
of $\mu$ and $\sigma^2$ are obtainable
from Eqs. (\ref{eq:P5}) and (\ref{eq:P6}),
we may apply our moment method to the FPE given by 
Eqs. (\ref{eq:A3}), (\ref{eq:P1}) and (\ref{eq:P2}) 
with the following steps:
(1) adopting the stationary PDF
given by Eqs. (\ref{eq:P3})-(\ref{eq:P4}), and
(2) expressing its parameters in terms of
the time-dependent $\mu(t)$ and $\sigma(t)^2$
in an appropriate way, as mentioned for models A, B and C.

As an application of our method, we have calculated 
the Fisher information for the dynamical inverse-gamma distribution,
which is realized for $\beta=0.0$ in the model A [Eq. (\ref{eq:E8})],
\begin{eqnarray}
p(x, t) &=& \frac{\kappa^{\delta(t)-1}}{\Gamma[\delta(t)-1]}
\:x^{-\delta(t)} e^{-\kappa(t)/x } \: \Theta(x),
\label{eq:N1}
\end{eqnarray}
with the time-dependent $\delta(t)$ and $\kappa(t)$
given by Eqs. (\ref{eq:E9}) and (\ref{eq:E10}). 
With the use of Eq. (\ref{eq:N1}), 
the Fisher information matrix given by
\begin{eqnarray}
g_{ij} &=& \left< \left( \frac{\partial \ln p(x)}{\partial \theta_i}\right) 
\left( \frac{\partial \ln p(x)}{\partial \theta_j} \right) \right>,
\label{eq:N2}
\end{eqnarray}
is expressed by
\begin{eqnarray}
g_{\delta\delta} &=& \psi'[\delta(t)-1], 
\label{eq:N3}\\
g_{\kappa \kappa} &=& \frac{[\delta(t)-1]^2}{\kappa(t)^2}, 
\label{eq:N4} \\
g_{\kappa \delta} &=&  \frac{[\delta(t)-1]}{\kappa(t)}
\{ \psi[\delta(t)-1]-\psi[\delta(t)] \},
\label{eq:N5}
\end{eqnarray}
where $\psi(x)$ and $\psi'(x)$ are  
di- and tri-gamma functions, respectively.

Figure \ref{figD} shows 
the time-dependent inverse-gamma distribution $p(x,t)$ 
when an input pulse given by Eq. (\ref{eq:M1}) 
with $\Delta I=0.3$ and $I_b=0.2$ is applied
to the model A with $\lambda=1.0$, $\alpha=0.5$ and $\beta=0.0$.
Solid and dashed curves show the results of the moment method
and the PDEM, respectively.
The time dependences of $\mu(t)$ and $\sigma(t)^2$
and of Fisher information are plotted
in Fig. \ref{figE}(a) and (b), respectively,
where solid (dashed) curves express
the result of the moment method (PDEM).
By an applied pulse at $2.0 \leq t < 6.0$, 
$\mu(t)$ and $\sigma(t)^2$ are increased and
the position of the PDF moves
rightward with an increased width.
An applied pulse increases $g_{\kappa \kappa}$
while it decreases $g_{\delta \kappa}$.
An interesting behavior is observed in $g_{\delta \delta}$
which is decreased at $t=2.0$ when a pulse is applied, 
but afterward it seems to gradually reduce to the stationary value.
A similar behavior is realized in $g_{\delta \delta}$ also at $t \geq 6.0$
when the applied pulse is off.

In our previous paper \cite{Hasegawa08a}, 
we applied the $q$-moment approach to the model A, 
deriving equations similar to Eqs. (\ref{eq:C6})-(\ref{eq:C9}) 
in the (normal) moment approach.
There are some differences between the $q$- and normal-moment approaches.
The stationary variance $\sigma_q^2$ in the $q$-moment approach
evaluated over the escort distribution is stable
for $0 \leq \alpha^2/\lambda < \infty $
whereas $\sigma^2$ in the normal-moment approach
is stable for $0 \leq \alpha^2/\lambda < 1.0 $
[Eqs. (\ref{eq:B9}), (\ref{eq:F8}) or (\ref{eq:H14})].
Although the time dependences of $\mu_q(t)$ and $\sigma(t)_q^2$ 
calculated by the $q$-moment approach are
similar to those of $\mu(t)$ and $\sigma(t)^2$ for a small 
$ \alpha $, the difference between them becomes significant 
for a large $ \alpha $ (see Fig. 13 in Ref. \cite{Hasegawa08a}). 
These differences yield the quantitative difference in $p(x,t)$ 
calculated by the normal- and $q$-moment methods, although both 
the methods lead to qualitatively similar results.
We note that in the limit of $\alpha=0.0$ ({\it i.e.,} additive noise only), 
the dynamical solution given by the $q$- or normal-moment method reduces to 
the Gaussian solution given by Eqs. (\ref{eq:C1})-(\ref{eq:C3}).
Thus the $q$- or normal-moment approach is a 
generalization of the Gaussian solution to the FPE 
with $\alpha \neq 0.0$ given by Eq. (\ref{eq:A3}).

In summary, by using the second-order moment method,
we have discussed the analytic time-dependent solution of
the FPE which includes additive and  multiplicative noise 
as well as external perturbations. It has been demonstrated that
dynamical PDFs calculated by the moment approach are in good agreement
with those obtained by the PDEM. Our moment method has some disadvantages.
The variance $\sigma^2$ diverges at $\alpha^2/\lambda \geq 1.0$,
for which our method cannot be applied.
If an applied perturbation induces a large $\mu(t)$ and/or $\sigma(t)^2$,
our method leads to poor results which are
not in good agreement to those calculated by the PDEM.
These are inherent in the moment approximation in which each moment
is required to be small.
Despite these disadvantages, however,
our moment method has following advantages: 
(i) obtained dynamical solutions are compatible with 
the exact stationary solutions in the Langevin models A, B and C,
(ii) it is useful for various subjects 
in which analytical dynamical PDFs are indispensable
({\it e.g.,} Ref. \cite{Hasegawa08a}), and 
(iii) the second-order moment approach is more tractable 
than sophisticated methods \cite{Paola02}-\cite{Attar08}
for the FPE subjected to multiplicative noise.
As for the item (iii), it is possible to take account of contributions 
from higher-order moments than second-order ones
with the use of Eq. (\ref{eq:A10}),
though actual calculations become tedious.

\section*{Acknowledgments}
This work is partly supported by
a Grant-in-Aid for Scientific Research from the Japanese 
Ministry of Education, Culture, Sports, Science and Technology.  

\newpage



\newpage

\begin{figure}
\begin{center}
\end{center}
\caption{
(Color online)
The time-dependent probability distribution
$p(x,t)$ of the model B with $\lambda=1.0$, $\alpha=0.5$, 
$\beta=0.5$ and $\epsilon=0.5$,
in response to an applied pulse given by Eq. (\ref{eq:M1})
with $\Delta I=0.5$ and $I_b=0.0$:
solid and dashed curves denote results
calculated by the moment method
and the PDEM, respectively,
curves being consecutively shifted downward
by 0.25 for a clarity of the figure.
}
\label{figK}
\end{figure}

\begin{figure}
\begin{center}
\end{center}
\caption{
(Color online)
The time dependence of $\mu(t)$ and 
$\sigma(t)^2$ of the model B with
$\lambda=1.0$, $\alpha=0.5$, $\beta=0.5$ and $\epsilon=0.5$,
calculated by the moment method 
with Eqs. (\ref{eq:J3}) and (\ref{eq:J4}) (solid curves)
and by the PDEM (dashed curves)
in response to an applied pulse $I(t)$ 
given by Eq. (\ref{eq:M1}) with $\Delta I=0.5$ and $I_b=0.0$
(the chain curve).
}
\label{figJ}
\end{figure}

\begin{figure}
\begin{center}
\end{center}
\caption{
(Color online)
The time-dependent probability distribution
$p(x,t)$ of the model C with
$\lambda=1.0$, $\alpha=0.5$, 
$\beta=0.5$, $r=0.5$ and $s=0.2$,
in response to an applied pulse given by Eq. (\ref{eq:M1})
with $\Delta I=0.5$ and $I_b=0.0$:
solid and dashed curves denote results
calculated by the moment method
and the PDEM, respectively,
curves being consecutively shifted downward
by 0.25 for a clarity of the figure.
}
\label{figB}
\end{figure}

\begin{figure}
\begin{center}
\end{center}
\caption{
(Color online)
The time dependence of $\mu(t)$ and 
$\sigma(t)^2$ of the model C with
$\lambda=1.0$, $\alpha=0.5$, $\beta=0.5$, $r=0.5$ and $s=0.2$,
calculated by the moment method 
with Eqs. (\ref{eq:J3}) and (\ref{eq:J4}) (solid curves)
and by the PDEM
(dashed curves),
in response to an applied pulse $I(t)$ 
given by Eq. (\ref{eq:M1}) with $\Delta I=0.5$ and $I_b=0.0$
(the chain curve).
}
\label{figA}
\end{figure}

\begin{figure}
\begin{center}
\end{center}
\caption{
(Color online)
The time-dependent probability distribution
$p(x,t)$ of the model C with $I=0.0$, $\alpha=0.5$, 
$\beta=0.0$, $r=0.5$ and $s=0.2$,
when the time-dependent relaxation rate $\lambda(t)$
given by Eq. (\ref{eq:M2}) is applied:
solid and dashed curves denote results
calculated by the moment method
and the PDEM, respectively,
curves being consecutively shifted downward
by 0.25 for a clarity of the figure.
}
\label{figF}
\end{figure}

\begin{figure}
\begin{center}
\end{center}
\caption{
(Color online)
The time dependence of $\mu(t)$ and 
$\sigma(t)^2$ of the model C with
$\lambda=1.0$, $\alpha=0.5$, $\beta=0.5$, $r=0.5$ and $s=0.2$,
calculated by the moment method
with Eqs. (\ref{eq:J3}) and (\ref{eq:J4}) (solid curves)
and by the PDEM
(dashed curves)
when the time-dependent relaxation rate $\lambda(t)$ 
given by Eq. (\ref{eq:M2}) is applied (the chain curve),
$\lambda$ being divided by a factor of five.
}
\label{figG}
\end{figure}

\begin{figure}
\begin{center}
\end{center}
\caption{
(Color online)
The time-dependence of the inverse-gamma distribution
$p(x,t)$ of the model A with $\lambda=1.0$, $\alpha=0.5$ 
and $\beta=0.0$
in response to an applied pulse given by Eq. (\ref{eq:M1})
with $\Delta I=0.3$ and $I_b=0.2$:
solid and dashed curves denote results calculated by the moment 
method and the PDEM, respectively.
Curves are consecutively shifted downward
by 1.0 for a clarity of the figure.
}
\label{figD}
\end{figure}

\begin{figure}
\begin{center}
\end{center}
\caption{
(Color online)
(a) The time dependence of $\mu(t)$ and $\sigma(t)^2$ 
of the model A 
with $\lambda=1.0$, $\alpha=0.5$ and $\beta=0.0$
calculated by the moment method (solid curves)
and PDEM (dashed curves)
in response to an applied pulse given by Eq. (\ref{eq:M1})
with $\Delta I=0.3$ and $I_b=0.2$  (the chain curve),
$\sigma(t)^2$ being multiplied by a factor of five.
(b) The time dependence of the Fisher information matrix
calculated by Eqs. (\ref{eq:N3})-(\ref{eq:N5}):
$g_{\delta \delta}$ (the solid curve), 
$g_{\kappa \kappa}$ (the dashed curve) and
$g_{\delta \kappa}$ (the chain curve), $g_{\delta \delta}$ and
$g_{\kappa \kappa}$ being multiplied 
by factors of 10 and 1/10, respectively (see text).
}
\label{figE}
\end{figure}
\end{document}